\documentclass[conference]{IEEEtran}
\IEEEoverridecommandlockouts
\usepackage[T1]{fontenc}
\usepackage{cite}
\usepackage{amsmath,amssymb}
\usepackage{graphicx}
\usepackage{booktabs}
\usepackage{xcolor}
\usepackage{tikz}
\usetikzlibrary{positioning,arrows.meta,calc}
\usepackage{url}
\usepackage[hidelinks,pdfauthor={Karan Yadav},pdftitle={Runtime BiDi Structural Verification Catches Right-to-Left Layout Defects that Accessibility, Actionability, and Logical-Property Linting Miss}]{hyperref}

\newcommand{\BulmaIssueRef}{\texttt{jgthms/bulma\#4063}}
\newcommand{\BulmaReportedSentence}{we reported it upstream as \texttt{jgthms/bulma\#4063}}
\newcommand{\AcknowledgmentText}{Claude (Anthropic) was used to assist with background research,
and portions of the tooling and prose were drafted with its assistance; the author
reviewed and revised all AI-assisted content, and all experiments were executed and
verified by the author. The artifact (detector, both modes, baselines, benchmarks, and
raw results) will be made publicly available. The Bulma \texttt{.menu-list} defect was
reported upstream (\texttt{jgthms/bulma\#4063}; no maintainer response at the time of writing).}

\begin{document}

\title{Runtime BiDi Structural Verification Catches Right-to-Left Layout Defects that Accessibility, Actionability, and Logical-Property Linting Miss}

\author{\IEEEauthorblockN{Karan Yadav}\IEEEauthorblockA{Google}}

\maketitle

\begin{abstract}
Bidirectional (right-to-left, RTL) layout defects
arise when interfaces authored with \emph{physical} directional CSS
(\texttt{margin-left}, \texttt{border-left}, \texttt{text-align:left}) are rendered under \texttt{dir="rtl"} without being mirrored. Accessibility
engines (axe-core), actionability engines (Playwright), and even
Stylelint's current core logical-property rule systematically fail to
localise these defects. We present runtime structural invariants---a \emph{zero-golden} mode over computed style and geometry, and
a bidi-conservative \emph{differential} mode---that catch them in a
headless browser. On a held-out benchmark of 14 real RTL bugs
reconstructed from dated fix commits in Bootstrap and Material
Components Web, the detector recovers 13/14 (2 via zero-golden,
11 via the differential mode), with no false alarm on any patched
version; on genuinely real historical compiled CSS it catches 1 of
2 attempted (the miss consistent with the reconstruction), using
the \emph{actual committed} \texttt{bootstrap.rtl.css}. On 16 negative
LTR-island cases (8 of them hard) it reports 0 false positives,
and both modes report 0 false positives on 43 unmodified real
components across four design systems and three RTL architectures (rule-of-three 95\% upper bound $\leq 7\%$). Along the way it
\emph{surfaced a real directional defect}---a physical \texttt{text-align:left}
in \texttt{.menu-list}---in the current Bulma 1.0.4. By contrast,
Stylelint's core rule emits an \emph{identical} 1744 warnings on Bootstrap's LTR and correct-RTL builds (it cannot triage a defect
from intentional physical CSS), pseudo-localisation detects none
of the mirroring bugs, and a naive pixel mirror-and-diff reaches
only F1 $0.87$ with a test-set-tuned threshold, whereas the differential needs none. We measure the two modes' limits directly
and find them \emph{complementary}: on a decoration/styling-asymmetry
class, zero-golden recovers 5/6 physical-left bugs but false-flags
the correct logical \texttt{border-inline-end} (4 of 4 border cases;
0 of 2 padding/margin), which computes to a left border under
RTL, while the differential detects 0/6 of the same bugs yet
never false-alarms---so on our benchmarks the differential is
\emph{empirically exact} (0 false alarms, no misses) only on the block-level inset/margin class, and zero-golden is the sole detector of
the decoration class. We scope this paper to detection; repair is
future work.
\end{abstract}


\section{Introduction}

Web interfaces are increasingly localised into RTL scripts
(Arabic, Hebrew, Persian, Urdu). The correct, direction-agnostic way to author them is with CSS \emph{logical} properties
(\texttt{margin-inline-start}, \texttt{border-inline-start},
\texttt{text-align:start}), which resolve to the physical right
edge under \texttt{dir="rtl"}. In practice, however, real design
systems remain overwhelmingly \emph{physical}. Measuring six
shipped stylesheets (Section~\ref{sec:prevalence}) we find 111--894 physical
directional declarations each and near-zero logical adoption
(the sole exception, Bulma, still ships 157); the dominant
coping strategy is to \emph{maintain a second, machine-mirrored
stylesheet} (e.g.\ Bootstrap ships both \texttt{bootstrap.css} and
\texttt{bootstrap.rtl.css}). A component rendered under
\texttt{dir="rtl"} with the un-mirrored stylesheet---a common
and easily made \emph{misconfiguration}---then exhibits directional
defects: start-side accents on the wrong edge, unmirrored
directional icons, and popups clipped at the wrong viewport
boundary.

We set out to build a \emph{zero-golden} detector---one needing
no reference render---and found empirically that the dominant
real-world defect class is not the styling asymmetry it targets
but physical \emph{inset/margin offsets}, which require a second,
\emph{differential} render: on real mined bugs the zero-golden mode
accounts for only 2 of 13 detections. The two modes prove
\emph{complementary}---zero-golden is the only mode that detects the
decoration/styling-asymmetry class, the differential the only
mode for the dominant inset/margin class---and that division
of labour is this paper's empirical arc.

Existing structural web-verification tools were not designed
for this failure mode. ReDeCheck~\cite{redecheck} detects element-overlap
and responsive-breakpoint faults; VizAssert~\cite{vizassert} verifies visual-logic assertions; and internationalisation-presentation tools~\cite{alameer2016,ifix}
model \emph{horizontal string expansion} between translations,
not directional mirroring. Industrial engines target orthogonal
properties: axe-core~\cite{axe} checks accessibility rules~\cite{wcag} and
Playwright~\cite{playwright} checks pointer actionability. Static linters can
flag physical properties in source, but---as we show---they
are directionally blind and cannot see inline, computed, or
runtime-injected styles.

\textbf{Contributions.} (1)~Runtime RTL structural invariants in
two \emph{complementary} modes---\emph{zero-golden} (single render, a
heuristic; the only mode that detects the decoration/styling-asymmetry class) and \emph{differential} (bidi-conservative; empirically exact on the block-level inset/margin class it dominates,
11 of 13 detections)---with the empirical finding that neither
mode subsumes the other. (2)~A held-out recall evaluation
on 14 real RTL bugs reconstructed from dated fix commits (two cross-checked against the actual committed historical \texttt{bootstrap.rtl.css}, one caught and one missed),
and a precision evaluation on 43 unmodified real library
components---all code we did \emph{not} author---together with 16
\emph{authored} LTR-island micro-benchmarks (8 genuinely hard)
that stress the RTL-script precision gate. (3)~A reproducible
comparison against \emph{real} axe-core, Playwright, Stylelint's current core rule, pseudo-localisation, and a naive pixel mirror-and-diff, plus a prevalence and RTL-bug-history study of six
shipped design systems. We scope this paper to \emph{detection};
repair is future work.

\section{Related Work and Positioning}

We position our detector against prior academic and industrial tooling. Prior i18n-presentation work~\cite{alameer2016,ifix} compares
a translated page against a baseline to find \emph{string-expansion}
overflow; it does not verify directional mirroring under
\texttt{dir="rtl"}. VizAssert~\cite{vizassert} and its successor~\cite{modular} verify formal
visual-logic assertions over layout. IFix~\cite{ifix} and XFix~\cite{xfix}
repair layout by constraint solving. ReDeCheck~\cite{redecheck} models
geometry and responsive breakpoints; Alameer et al.'s i18n
tooling~\cite{alameer2016} uses a layout-graph model for string-expansion
failures, not directional structure. axe-core and Playwright are
industrial engines for accessibility and actionability respectively. None target RTL physical-vs-logical structure. Text-level bidi checkers (e.g.\ Google's BidiChecker) verify Unicode Bidirectional Algorithm conformance and mixed-script
transposition within strings; our concern is orthogonal---CSS
box-model geometry and physical-vs-logical layout mirroring.
Two practitioner baselines are relevant: \emph{pseudo-localisation}
(accented, lengthened, RTL-wrapped strings) surfaces truncation and expansion but not physical-vs-logical mirroring, and
\emph{mirror-and-diff} visual regression (flip the LTR screenshot, diff
against RTL) is the naive pixel competitor to our differential
mode (Section~\ref{sec:baselines}); DOM-aware structural visual-regression
tools---commercial or academic---filter pixel noise but remain
direction-agnostic, encoding no start/end semantics, so they
too cannot triage a mirroring defect without the logical-vs-physical knowledge our invariants add.

\section{RTL Structural Invariants}
\label{sec:invariants}

We render each state in real headless Chromium and evaluate invariants over the computed style map and bounding boxes
of every element whose computed \texttt{direction} is \texttt{rtl}. The
invariants are \emph{zero-golden}: they need no LTR reference rendering. Zero-golden detection is a \emph{heuristic} that assumes \emph{start-side} decoration intent: a logical start-side decoration resolves
to the physical right edge under \texttt{dir="rtl"}, so a decoration
still on the physical left edge is \emph{evidence}---not proof---that a
physical property was used where a logical one was intended.
Throughout, we call a mode \emph{empirically exact} on a defect class
when, on our benchmarks, it produces no false alarms and
no misses---an empirical property on the classes we test, not
a formal soundness guarantee. Zero-golden cannot be made
exact in this sense: per CSS, logical properties \emph{compute to}
their physical counterparts, so \texttt{getComputedStyle} cannot
distinguish an authored \texttt{border-left} from a correct logical
\texttt{border-inline-end}, which under RTL also computes
to a left border. We measure this false-positive class in
Section~\ref{sec:fp}; there the differential never false-alarms on it,
but---as we also measure---never \emph{detects} this class either, so
the two modes are complementary rather than one dominating.

\begin{itemize}
\item \textbf{Physical inline asymmetry}: a left-only visible
border, a left-dominant padding asymmetry,
or \texttt{text-align:left} on RTL running text
(\texttt{RTL\_PHYSICAL\_BORDER\_/\allowbreak PADDING\_/\allowbreak TEXT\_ALIGN}). These are gated on the element
containing RTL-script (Arabic/Hebrew) text, so legitimate
left-aligned LTR islands (English or numeric columns,
code, \texttt{<bdi>} phone/email) are not flagged.
\item \textbf{Viewport clipping}: an absolutely/fixed-positioned popup,
narrower than the viewport, whose box overflows
$[0, w_{\mathrm{viewport}}]$ under RTL (\texttt{RTL\_VIEWPORT\_CLIP}).
\item \textbf{Directional-icon mirroring}: a directional glyph (arrow,
chevron, send) not horizontally reflected (\texttt{scaleX(-1)})
under RTL (\texttt{RTL\_ICON\_NOT\_MIRRORED}). Unlike the
other invariants this one needs a per-icon directional
annotation, so we exercise it only on the controlled
benchmark (Table~\ref{tab:controlled}) and exclude it from every real-code
evaluation, where icons are unannotated.
\end{itemize}

Because the invariants read \emph{computed} style, they detect
defects regardless of whether the offending property originates in a stylesheet, an inline \texttt{style} attribute, or runtime
JavaScript---a reach no static linter has.

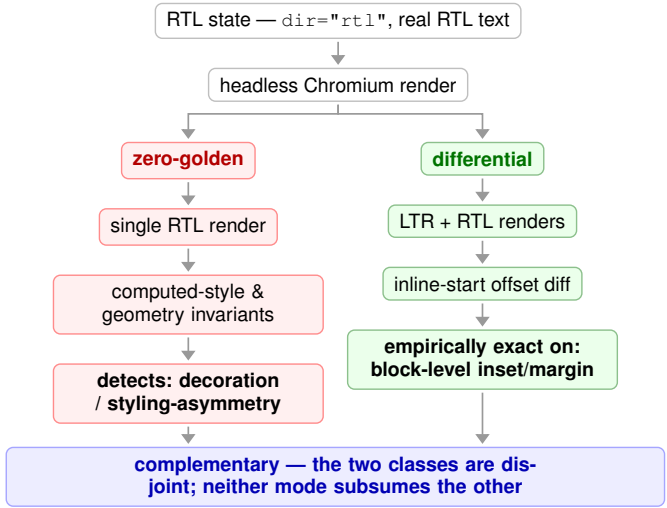
\begin{figure}[t]
\centering
\resizebox{\columnwidth}{!}{\begin{tikzpicture}[
  font=\sffamily\scriptsize,
  >={Triangle[width=5pt,length=3.5pt]},
  box/.style={draw=black!25, rounded corners=2.5pt, fill=white, align=center,
              inner xsep=4pt, inner ysep=3.2pt, line width=0.5pt},
  zg/.style={box, draw=red!45, fill=red!6},
  df/.style={box, draw=green!45!black!45, fill=green!8},
  cmp/.style={box, draw=blue!35, fill=blue!7},
  arr/.style={->, black!40, line width=0.6pt, shorten >=1pt, shorten <=1pt},
  node distance=3.6mm
]
  \node[box] (state) {RTL state --- \texttt{dir="rtl"}, real RTL text};
  \node[box, below=of state] (render) {headless Chromium render};

  \node[zg, below left=5mm and -6mm of render, text=red!70!black, font=\sffamily\scriptsize\bfseries] (zg) {zero-golden};
  \node[df, below right=5mm and -6mm of render, text=green!45!black, font=\sffamily\scriptsize\bfseries] (df) {differential};

  \node[zg, below=of zg] (zg1) {single RTL render};
  \node[zg, below=of zg1, text width=31mm] (zg2) {computed-style \& geometry invariants};
  \node[zg, below=of zg2, text width=31mm, font=\sffamily\scriptsize\bfseries] (zg3) {detects: decoration / styling-asymmetry};

  \node[df, below=of df] (df1) {LTR + RTL renders};
  \node[df, below=of df1] (df2) {inline-start offset diff};
  \node[df, below=of df2, text width=31mm, font=\sffamily\scriptsize\bfseries] (df3) {empirically exact on: block-level inset/margin};

  \path (zg3.south) -- (df3.south) coordinate[midway] (mid);
  \node[cmp, below=5mm of mid, text=blue!70!black, font=\sffamily\scriptsize\bfseries, text width=0.9\columnwidth] (cmp)
       {complementary --- the two classes are disjoint; neither mode subsumes the other};

  \draw[arr] (state) -- (render);
  \draw[arr] (render.south) -- ++(0,-1.5mm) -| (zg.north);
  \draw[arr] (render.south) -- ++(0,-1.5mm) -| (df.north);
  \draw[arr] (zg) -- (zg1);
  \draw[arr] (zg1) -- (zg2);
  \draw[arr] (zg2) -- (zg3);
  \draw[arr] (df) -- (df1);
  \draw[arr] (df1) -- (df2);
  \draw[arr] (df2) -- (df3);
  \draw[arr] (zg3.south) -- (zg3.south |- cmp.north);
  \draw[arr] (df3.south) -- (df3.south |- cmp.north);
\end{tikzpicture}}
\caption{The two runtime modes. \emph{Zero-golden} needs a single RTL render
and detects the decoration/styling-asymmetry class (its precision limit is a
\texttt{border-inline-end} false positive); the \emph{differential} needs an LTR and
an RTL render and is empirically exact on the block-level inset/margin class.
The classes are disjoint, so the modes are complementary (Sections~\ref{sec:fp}
and~\ref{sec:complementarity}).}
\label{fig:modes}
\end{figure}

Physical inset/margin offsets (e.g.\ \texttt{right:0} vs.\
\texttt{inset-inline-start:0}) resolve to the same used
value in a single render and are therefore invisible to
zero-golden detection. For these we add a \emph{differential}
mode (summarised in Figure~\ref{fig:modes}): we render the \emph{identical}
component---same DOM and text---under \texttt{dir=ltr} and
\texttt{dir=rtl} (a directional toggle, not a cross-lingual translation
pair) and flag any block-level element whose distance from
the inline-start edge of its parent fails to mirror. Unlike
the zero-golden invariants, the differential is not gated on
script---it compares \emph{every} block-level element, so an LTR
island (a code block or numeric table) embedded in an RTL
layout is still checked, since its placement is governed by the
RTL parent and must mirror. We call it \emph{bidi-conservative}: it
skips inline-level elements, because Unicode bidi legitimately
reorders them across directions---avoiding that false-positive
class at the cost of inline-margin defects (Section~\ref{sec:recall}).
The RTL-script gate uses the Arabic and Hebrew Unicode
ranges and inspects an element's \emph{descendant} text, so a defect
on a wrapper whose text lives in a child is still covered;
but an element with no RTL text anywhere is not, so the
detector must be run against pages with real RTL content,
not placeholder text.

\textbf{Parameters.} All thresholds are fixed constants, not tuned
per subject. The differential takes each block-level element's
distance from its parent's inline-start edge---the element's
left minus the parent's left under LTR, and the parent's
right minus the element's right under RTL---and flags a
mismatch above a \textbf{2\,px} tolerance (absorbing sub-pixel rounding); it ignores boxes under $4{\times}4$\,px and elements whose
width changes by more than 4\,px between directions (reflow). ``Inline-level'' means computed \texttt{display:inline}
only, so \texttt{inline-block}, \texttt{-flex}, and \texttt{-grid} boxes \emph{are}
compared. Zero-golden's padding-asymmetry rule fires only
when left padding exceeds right by more than 4\,px \emph{and}
the element also carries \texttt{text-align:left} or a left-only visible border; the viewport-clip rule ignores full-bleed
boxes (width $\geq 0.9\,w_{\mathrm{viewport}}$) and flags any box exiting $[-0.5, w_{\mathrm{viewport}}{+}0.5]$ after skipping elements that are
\texttt{display:none}, \texttt{visibility:hidden}, or fully transparent. Because only the \texttt{dir} attribute changes between the
two differential renders they share an identical DOM, so
elements are matched by document order, and ``parent'' denotes
the immediate DOM parent (a simplification for absolutely-positioned boxes whose containing block differs).

\section{Evaluation}
\label{sec:eval}

\textbf{Setup.} All results are from real execution on
one VM: Node~22.14, Chromium/Headless-Shell
153.0.8010.12 (Playwright 1.63.0), axe-core 4.13.0
via \texttt{@axe-core/playwright} 4.13.0, and
Stylelint 17.15.0~\cite{stylelint} using its current core layout-mappings rules (\texttt{property-layout-mappings} and
\texttt{value-keyword-layout-mappings}, primary option
\texttt{flow-relative}). Renders use the fonts bundled with the
pinned Playwright Chromium build; we install no additional
system fonts, so glyph metrics are fixed by that build.

\subsection{Prevalence of physical directional CSS}
\label{sec:prevalence}

Table~\ref{tab:prevalence} counts directional declarations in six shipped
stylesheets. We count declarations by matching each property
name at a declaration boundary in the CSS text (``phys.\ box-dir'' $=$ \texttt{margin/\allowbreak padding/\allowbreak border-left|right} plus
\texttt{left|right} insets; ``t-align L/R'' $=$ \texttt{text-align:left|}\allowbreak\texttt{right} values). Physical directional properties dominate and
logical adoption is negligible. Notably, \texttt{bootstrap.css}
and \texttt{bootstrap.rtl.css} carry the \emph{same} 644 physical
declarations: RTLCSS rewrites direction (\texttt{-left}$\leftrightarrow$\texttt{-right})
without reducing physicality---the mirrored build is just as
physical, which is precisely why a second stylesheet must
be maintained. Bulma is the sole logical-property-first library, yet still ships 157 physical declarations and a physical
\texttt{text-align:left} that breaks its menu under RTL (Section~\ref{sec:fp}).

\begin{table}[t]
\caption{Directional declarations in real shipped stylesheets.}
\label{tab:prevalence}
\centering
\begin{tabular}{@{}lrrr@{}}
\toprule
Stylesheet & phys.\ box-dir & logical & t-align L/R \\
\midrule
bootstrap.css (LTR) & 644 & 6 & 18 \\
bootstrap.rtl.css (RTL) & 644 & 6 & 18 \\
primer.css & 894 & 3 & 18 \\
material-components-web.css & 111 & 0 & 4 \\
foundation.css & 334 & 0 & 13 \\
bulma.css & 157 & 217 & 22 \\
\bottomrule
\end{tabular}
\end{table}

\subsection{Static linting cannot triage directional defects}

Stylelint's current core rule \texttt{property-layout-mappings} (\texttt{flow-relative})
is a source-level \emph{convention} rule: it flags physical properties
that have flow-relative equivalents. On Bootstrap's LTR build
it emits 1744 warnings; on Bootstrap's \emph{correct} RTL build it
emits the \emph{identical} 1744 (Primer 2304, MDC 2481, Bulma
791). Emitting 1744 warnings is the \emph{correct} behaviour for
a convention rule---but the count is unchanged between
a possibly-buggy sheet and a known-correct one, so the
output cannot be triaged into defects; and a convention rule
fundamentally cannot see inline, computed, cascaded, or
runtime-injected styles. This is a scope mismatch (convention
checker $\neq$ defect detector) plus an un-triageable volume,
not a fault in Stylelint. (Direction-aware linter plugins add
scoping directives such as \texttt{/*rtl:ignore*/}, but share the
fundamental limit that no source linter sees computed, inline,
cascaded, or JS-injected styles.)

\subsection{Controlled benchmark}

We built 12 self-contained states (6 buggy, 6 clean) exercising the invariant classes, with the defect placed in a stylesheet,
an inline attribute, or runtime-injected JS. Table~\ref{tab:controlled} reports
real detection performance; axe-core and Playwright appear as
\emph{negative controls}---standard accessibility and actionability CI
gates that target orthogonal specifications by design---rather
than competing layout detectors, so their low recall shows
only that passing those gates gives false confidence about RTL
layout. axe-core detects none of the RTL defects (it ships no
directional rule); Playwright detects only the geometric clip
and false-positives on an \texttt{sr-only} region; Stylelint detects
only the two stylesheet-located defects and misses the inline,
injected, icon, and geometric ones. We note plainly that these
states were authored by us to exercise the invariants; the next
study uses unmodified third-party code.

\begin{table}[t]
\caption{Detection on the controlled benchmark (n=12; positive=buggy).}
\label{tab:controlled}
\centering
\setlength{\tabcolsep}{4.5pt}
\begin{tabular}{@{}lrrrrrrr@{}}
\toprule
Tool & TP & FP & TN & FN & P & R & F1 \\
\midrule
Detector (ours) & 6 & 0 & 6 & 0 & 100 & 100 & 100 \\
axe-core 4.13.0 & 0 & 0 & 6 & 6 & -- & 0 & -- \\
Playwright & 1 & 1 & 5 & 5 & 50 & 17 & 25 \\
Stylelint (core) & 2 & 0 & 6 & 4 & 100 & 33 & 50 \\
\bottomrule
\end{tabular}
\end{table}

\subsection{False positives on unmodified real components}
\label{sec:fp}

To measure precision without authoring the code, we render \textbf{43 real components} under \texttt{dir="rtl"} using each
library's \emph{correct} RTL configuration, each populated with real
Arabic/Hebrew text (labels, headings, body) so the RTL-script gate is genuinely exercised rather than vacuously satisfied by English placeholder text; any flag is then a false
positive. We cover four design systems spanning the three
RTL architectures found in the wild: mirror stylesheet (Bootstrap 5.3.8, Foundation 6.9.0), \texttt{[dir=rtl]} overrides (Material Components Web 14; 526 override rules), and native
logical properties (Bulma 1.0.4). Correctness is defined per
architecture: the \texttt{.rtl.css} build (Bootstrap, Foundation);
the intact stylesheet vs.\ a \emph{mutation} that strips its \texttt{[dir=rtl]}
rules with PostCSS (MDC), i.e.\ mutation-based evaluation;
and the single logical-property stylesheet (Bulma). The detector reports \textbf{0 false positives across all 43 components}
(Table~\ref{tab:fp}); re-rendering each at 360/768/1280\,px (129 renders)
leaves this unchanged, so viewport is a robustness check,
not independent replication. With 0 events in $n = 43$, the
rule-of-three 95\% upper bound on the false-positive rate is
$3/43 \approx 7\%$. We separately run the \emph{differential} mode over the
same 43 correctly-built components (rendering each library's
real LTR and RTL configurations and checking that they
mirror): it too reports \textbf{0 false positives}, so both modes are
clean on real code. This 0/43 is consistent with zero-golden's
4/4 end-side false-positive rate below: zero-golden's pass \emph{is} the
runtime check for a left-only border on an RTL-text element
(invariant \texttt{RTL\_PHYSICAL\_BORDER\_ASYMMETRY}), and it
fired on none of the 43---so no correctly-built component
actually \emph{renders} such an element, and the end-side idiom
that defeats zero-golden does not arise in this real-component
sample. (Consistently, the only end-side \emph{border} rule we find
in the sampled libraries is Bulma's \texttt{.card-footer-item}
divider, absent from the sampled components.)

\begin{table}[t]
\caption{False positives on correctly-rendered real components, 3 viewports each.}
\label{tab:fp}
\centering
\begin{tabular}{@{}llrr@{}}
\toprule
Library & RTL architecture & comp. & FP \\
\midrule
Bootstrap 5.3.8 & mirror stylesheet & 14 & 0 \\
Foundation 6.9.0 & mirror stylesheet & 9 & 0 \\
MDC 14 & \texttt{[dir=rtl]} overrides & 10 & 0 \\
Bulma 1.0.4 & logical properties & 10 & 0$^{*}$ \\
\midrule
Total (components) & & 43 & 0 \\
\bottomrule
\end{tabular}

\vspace{2pt}
\parbox{\columnwidth}{\scriptsize $^{*}$We count the Bulma menu flag as a real bug (\BulmaIssueRef), not a false positive, giving 0/43; under a strict protocol that treats \emph{any} flag
on a shipped configuration as a false positive it is 1/43 ($\approx 2.3\%$;
rule-of-three upper bound $\approx 9\%$). Each component is also rendered at
360/768/1280\,px (129 renders) with no change; the differential mode is
likewise 0 FP on the 43.}
\end{table}

The one apparent exception is \emph{not} a false positive but a
\emph{real bug the detector found}: Bulma 1.0.4---though logical-property-first (157 physical vs.\ 217 logical declarations)---still
ships \texttt{.menu-list a \{ text-align: left \}}, so its
menu text hugs the wrong edge under \texttt{dir="rtl"}.
Under the un-adapted condition the detector also flags
genuine defects, e.g.\ the shipped Bootstrap \texttt{.text-start}
utility (\texttt{text-align:left !important}) and left-aligned MDC text-field/data-table content that MDC's own
\texttt{[dir=rtl]} rules exist to fix. (Bulma does not advertise
explicit RTL support; its logical properties provide it
implicitly, so the menu flag is a latent directional defect rather
than a regression against a promised feature; \BulmaReportedSentence.)

\emph{Negative cases.} A clean set that merely removes left-side styling would make ``0 false positives'' near-tautological.
We therefore add \textbf{16 negative cases} (8 of them genuinely
hard) on which any flag would be a false positive: English
SKU and numeric table columns, URLs, an LTR code block
(including one with a physical left accent border), inline
\texttt{<code>}, phone/email in \texttt{<bdi>}, an isolated LTR identifier,
non-mirroring icons (checkmark, search), a brand logo, an
English note, a progress bar, and a range scrubber. The detector
flags \textbf{0/16}, because the physical-decoration invariants are gated
on the element containing RTL-script text (Section~\ref{sec:invariants}); three
positive controls---RTL text with the same physical property---are all caught, confirming the gate does not cost recall \emph{on these
three controls}; the gate does forgo recall on elements with
no RTL text anywhere (icon-only controls, empty containers),
a deliberate precision trade and a deployment precondition
(Section~\ref{sec:threats}). By exclusion mechanism the 16 split into \textbf{8 whose
physical-left styling is suppressed only by the RTL-script
gate} (SKU/numeric columns, URLs, code, brand logo, English
note, mixed table) and 8 with no physical-left styling or an
annotation-gated icon; the gate---the load-bearing rule---is thus
exercised by 8, and the end-side cases below are the negatives
that actually stress the invariant logic.

\begin{figure}[t]
\centering
\includegraphics[width=0.98\columnwidth]{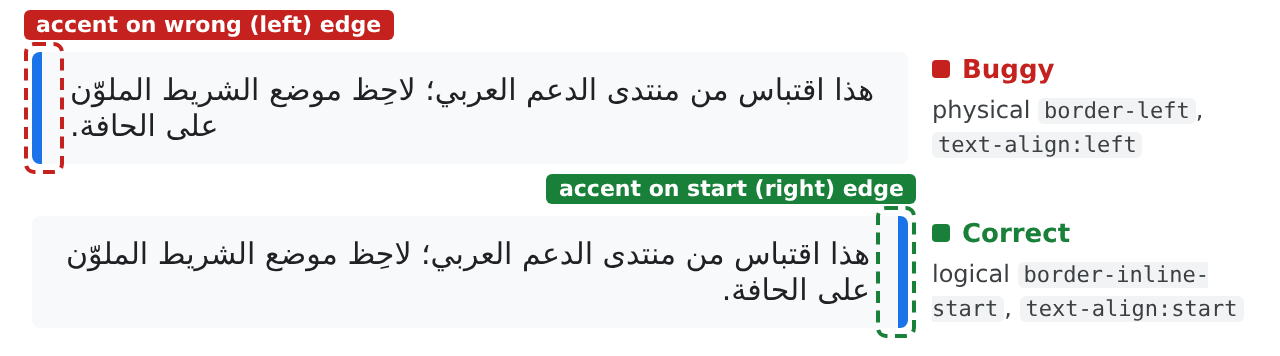}
\caption{An RTL blockquote. Top: a physical
\texttt{border-left}/\texttt{text-align:left} accent renders on the
wrong (left) edge---the detector flags it. Bottom: the logical
\texttt{border-inline-start}/\texttt{text-align:start} version renders on
the correct (right/start) edge and is not flagged.}
\label{fig:blockquote}
\end{figure}

The genuinely hard case is an element that \emph{does} contain RTL text and \emph{correctly} carries left-side styling. Under RTL a correct logical \texttt{border-inline-end} computes to a left border, indistinguishable from a physical
\texttt{border-left} by computed style. We measure this on 6
correct end-side decorations (4 \texttt{border-inline-end}, one
\texttt{padding-inline-end}, one \texttt{margin-inline-end}) on
RTL-text elements: the zero-golden heuristic \emph{false-positives on
4/4} of the \texttt{border-inline-end} cases (the class it cannot
distinguish) and none of the padding/margin cases, while the
differential mode is \emph{clean on all 6}. This is the precision limit
of the zero-golden heuristic; the same class turns out to be
the one only zero-golden can \emph{detect} (Section~\ref{sec:complementarity}). Figure~\ref{fig:blockquote}
shows a representative defect and its logical fix.

The MDC study additionally \emph{found a bug in our own
detector}: an early viewport-clip invariant fired on MDC's full-width fixed top-app-bar; we refined it to ignore full-bleed
elements (width $\geq 0.9\,w_{\mathrm{viewport}}$), removing the false positive
while preserving detection of narrow clipped popups. This full-bleed rule is a semantic invariant---a near-full-width bar cannot
be a clipped \emph{narrow} popup---rather than an MDC-specific
patch; but because it was prompted by the MDC set, the clip
invariant's MDC figure is not fully held-out (Section~\ref{sec:threats}).

\subsection{Mode complementarity: neither mode subsumes the other}
\label{sec:complementarity}

The precision limit above has a recall counterpart that makes
the two modes \emph{complementary} rather than one dominating. We
construct the mirror image of the end-side set---6 \emph{physical}-left decoration bugs (\texttt{border-left}, \texttt{text-align:left},
\texttt{padding-left}) on RTL text, the actual mirroring defect
those logical properties fix---and run both modes on this
constructed decoration set. Zero-golden recovers \textbf{5/6} and the
differential \textbf{0/6}; both are clean on the correct counterparts.
The single zero-golden miss is the \texttt{padding-left} case:
the padding-asymmetry rule fires only when left padding
exceeds right by more than 4\,px \emph{and} the element also carries \texttt{text-align:left} or a left-only border---a deliberately conservative co-condition---so a lone \texttt{padding-left}
(which has neither co-signal) is not flagged, whereas its
\texttt{border-left} and \texttt{text-align} siblings trip categorical
invariants directly. The differential's 0/6 is not a tuning artifact
but \emph{structural}: its only signal is an element's box offset
from the parent's inline-start edge, and a \texttt{border-left},
\texttt{text-align:left}, or \texttt{padding-left} does not move
that box, so the whole decoration/styling-asymmetry class is
invisible to it by construction---its zero false alarms there
come from never firing, not from soundness. Zero-golden
is thus the \emph{only} mode that detects this class, paying the
\texttt{border-inline-end} false positive of Section~\ref{sec:fp} as
the price, while the differential is empirically exact (0 false
alarms, no misses) on the disjoint block-level inset/margin
class it dominates. This division of labour is not an artifact
of constructed fixtures: on the 14 held-out real bugs reconstructed from Bootstrap and MDC fix commits (Section~\ref{sec:recall},
Table~\ref{tab:recall}), the partition is \emph{exact}---zero-golden catches 2 (both
text-align), the differential 11 (inset/margin), 1 is missed by
both, and \emph{no bug is caught by both modes}---so complementarity holds on code we did not write.

\subsection{Recall on real mined bugs}
\label{sec:recall}

The controlled benchmark measures recall only on bugs we
wrote. To measure recall on bugs we did \emph{not} write, we started
from the 53 mined RTL-fix commits (Section~\ref{sec:history}), kept the
24 whose diff changes a directional CSS declaration (19 MDC,
5 Bootstrap), and reconstructed the 14 that map to a single
renderable declaration into minimal held-out states (pre-fix
buggy, post-fix correct). Selection was by \emph{renderability of the
changed declaration}, decided before running the detector; no
commit was dropped after observing detector behaviour. The
excluded commits were SCSS-mixin/variable changes with
no single renderable declaration, or docs/build commits. By
bug class the 24 split as inset/offset 10, margin 6, text-align
2, border-radius 2, and other 4, and are MDC-heavy (19 of
24); the inset/margin-offset class dominates, which is why the
differential carries recall. The 14 reconstructions follow this
shape (11 MDC, 3 Bootstrap; 11 offset/margin, 2 text-align, 1
inline-margin miss)---so per-library recall cannot be estimated
from Bootstrap's $n = 3$.

We run both modes and count a bug as caught if either
fires; the correct version must pass both. The detector recovers
\textbf{13/14} (Table~\ref{tab:recall}), but the split is the real finding: only \textbf{2}
are caught by zero-golden (the styling-asymmetry class, e.g.\
MDC \texttt{bd8d1aaf} data-table alignment) and \textbf{11} by the differential mode (the inset/margin-offset class, which dominates
the mined fix commits). Both zero-golden catches are text-align styling-asymmetry bugs---both instances of its class in
this corpus ($n = 2$, both MDC; on Bootstrap zero-golden is
0/3)---while the differential is 11/11 on the dominant block-level inset/margin class, and no bug is caught by both modes---the exact partition that corroborates the mode complementarity
of Section~\ref{sec:complementarity}. The miss (\texttt{489e7b679}, an inline card-link
\texttt{margin-left}) is skipped by the bidi-conservative differential. The differential fires on \textbf{0/14 patched versions}. (Cost is
secondary: zero-golden's single render is a median \textbf{55\,ms} per
component and the differential's two renders \textbf{110\,ms} (IQR 54--56 vs.\ 108--113\,ms), so the second render roughly doubles the
cost---but the reason to keep zero-golden is coverage of the
decoration class, plus that it needs only one direction when
auditing a deployed RTL build one cannot re-render.) Because
these are minimal reconstructions, 13/14 is an \emph{upper bound}
on field recall (real pages add cascade, specificity, and sibling
noise).

\begin{table}[t]
\caption{Held-out recall on real RTL bugs reconstructed from dated fix commits.}
\label{tab:recall}
\centering
\begin{tabular}{@{}lrrr@{}}
\toprule
 & Bootstrap & MDC & Total \\
\midrule
reconstructed bugs & 3 & 11 & 14 \\
caught by zero-golden & 0 & 2 & 2 \\
caught by differential & 2 & 9 & 11 \\
missed & 1 & 0 & 1 \\
false alarm on patched version & 0 & 0 & 0 \\
\bottomrule
\end{tabular}

\vspace{2pt}
\parbox{\columnwidth}{\scriptsize Overall recall 13/14. The library split (MDC 11, Bootstrap 2) and the mode
split (differential 11, zero-golden 2) are \emph{different} partitions: both zero-golden
catches are MDC, so Bootstrap's two caught bugs are both differential. The
14 commits span 2017--2021, a recent subset of the 2014--2022 fix history.}
\end{table}

To validate on genuinely real code, we also render two of
these Bootstrap defects using the \emph{actual committed historical}
\texttt{bootstrap.rtl.css} from the buggy revision (Bootstrap
ships a single compiled RTL stylesheet, so the historical
build is directly recoverable; MDC's \texttt{[dir=rtl]} overrides
are interleaved in its main sheet and were not as cleanly
reconstructable, which is why the two spot-checks are Bootstrap's). The dropdown-menu-end defect (\texttt{57205e3d8}, real
\texttt{right:0;left:auto}) is caught by the differential on the
real compiled CSS; the inline card-link is missed, consistent
with the reconstruction.

\subsection{RTL bugs recur across a library's lifetime}
\label{sec:history}

We mined the full history (blobless clones) of each
repository with a case-insensitive commit-message query
(\texttt{fix.*rtl} or \texttt{rtl.*(bug|fix|incorrect|broken|}\allowbreak\texttt{wrong)}) and kept commits that touch \texttt{.scss}/\texttt{.css} with a
directional declaration; the 14 reconstructions of Section~\ref{sec:recall}
are drawn from this set. This yields 19 explicit RTL fixes
in Bootstrap and 34 in Material Components Web, spanning
2014--2022 and 2016--2022. Nineteen fixes over eight years
is a small fraction of Bootstrap's commit volume, so we do
not claim RTL is a dominant cost; the defensible observation
is that RTL layout defects \emph{recur across the entire lifetime} of
both libraries, including years after official RTL support was
added. Primer has 0 such fixes, consistent with its shipping
no RTL support at all (0 \texttt{[dir=rtl]} rules) rather than with
correctness.

\subsection{Baselines: pseudo-localisation and mirror-and-diff}
\label{sec:baselines}

We run the two practitioner baselines over the 14 held-out bugs. \emph{Pseudo-localisation} (RTL-wrap the text with isolate
marks, expand it, and bracket it, then check for overflow)
detects the directional defect in \textbf{0/14} (0 recall) and moreover
raises \textbf{2 false alarms}: it targets truncation and expansion, and
mirroring produces no overflow, while 2/14 show incidental
expansion overflow that is present on the correct versions
too. \emph{Mirror-and-diff} (screenshot the component---not the full
viewport---under LTR, flip it horizontally, and diff against
RTL with \texttt{pixelmatch}) cannot cleanly separate buggy from
correct: across the 14, the minimum buggy mismatch (0.2\%)
is \emph{below} the maximum correct mismatch (0.6\%), so the classes
overlap; its best achievable F1 is \textbf{0.87} (accuracy 0.86) and only
with a threshold tuned on the test set itself. The geometry
differential needs no threshold, reports exactly 0 on all 14
correct versions, and flags 13/14 buggy. We compare against
the standard pixel form of mirror-and-diff; a text-masking or
box-only variant would reduce its noise but is itself a step
toward the geometric comparison the differential performs
directly.

\section{Threats to Validity}
\label{sec:threats}

\textbf{Construct.} The controlled benchmark, the end-side false-positive set, and the decoration-recall set were authored by
us; we therefore de-emphasise their figures, rely on the
held-out recall (13/14) and negative-case precision (0/16,
8 hard) for generalisation, and report 13/14 as an \emph{upper
bound} on field recall. The mode complementarity is likewise
corroborated on the held-out real bugs, where the partition
is exact (Section~\ref{sec:complementarity}). \textbf{Complementary modes, not one
exact mode.} Zero-golden false-positives on correct end-side
(\texttt{border-inline-end}) decorations, which compute to a
left border under RTL; the differential does not false-alarm
there but also does not detect the decoration class at all (0/6),
so it is \emph{empirically exact} only on the block-level inset/margin
class (11/11, 0~FP) and the two modes are complementary
(Section~\ref{sec:fp}). \textbf{Precision generality.} The 0/43 field figure
reflects the physical-dominant prevalence we measured; on
codebases that heavily use native logical end-side properties (\texttt{border-inline-end} on RTL-text elements), zero-golden's false-positive rate rises toward the 4/4 we measure
on that idiom---the differential stays clean. \textbf{Tuning disclosure.} One invariant, the viewport clip, was refined after
a false positive on the MDC set (full-bleed suppression),
so its MDC 0-FP figure is not held-out; the refinement is
a general semantic rule and the other invariants were unchanged. \textbf{Baseline scope.} We benchmark the practitioner-default \emph{pixel} mirror-and-diff; DOM-aware structural visual-regression tools reduce pixel noise but are direction-agnostic
(no start/end semantics), so detecting RTL mirroring with them
needs exactly the logical-vs-physical knowledge our invariants
add. \textbf{Benchmark availability.} No public benchmark of RTL
directional-mirroring defects exists---responsive-layout~\cite{redecheck} and
i18n string-expansion~\cite{alameer2016} datasets target different classes---so
we mine and reconstruct our own; a full-page RTL corpus
is future work. \textbf{RTL-content precondition.} The RTL-script
gate means the detector finds nothing on pages rendered with
placeholder or English text, nor on textless controls (icon-only
buttons, empty inputs, pagination arrows) unless RTL text lives
in a descendant; it must be run against real RTL content, and
textless controls need explicit directional annotation. \textbf{Composition.} Components are rendered standalone; a component
embedded in a real page may inherit direction and context
we do not model, and minimal reconstructions omit cascade,
specificity, and sibling noise. \textbf{Rendering.} The detector's algorithm is deterministic, but headless layout depends on font
metrics; we pin the Playwright Chromium build and install
no extra fonts, so metrics are fixed but not identical to other
environments. \textbf{Coverage.} Inline-margin, float, and physical
corner-radius (\texttt{border-radius}) directional bugs are missed
(e.g.\ \texttt{489e7b679}), and a lone physical padding on a text-only
leaf block escapes both modes (zero-golden lacks its co-signal;
the differential's box does not move and its text children are
inline). The viewport-clip rule can in principle false-positive
on an off-canvas panel parked off-screen while still \texttt{visible};
the differential's 4\,px reflow filter can mask a defect that
itself shifts an element's width; and because the differential
compares \texttt{inline-block}/\texttt{-flex} boxes, one legitimately
bidi-reordered among mixed-script text could in principle
mismatch (none did in our sample). \textbf{External.} We evaluate
Arabic/Hebrew RTL; other scripts (Thaana, Syriac, N'Ko,
Adlam) and mixed bidi runs are untested. \textbf{Reconstruction.}
The held-out bugs are minimal reconstructions (a form of
mutation~\cite{mutation}) of the changed declarations, validated on two
cases with the real committed historical build.

\section{Conclusion}

Runtime BiDi structural verification detects RTL directional
defects that accessibility engines, actionability engines, and
Stylelint's core logical-property rule miss. On real bugs it did
not author it recovers 13/14 (2 via zero-golden, 11 via the
differential mode), with no false alarm on the patched code;
on legitimate LTR islands and 43 real components both modes
report 0 false positives; and it surfaced a real directional defect
in the current Bulma. The two modes are \emph{complementary}: the
differential is empirically exact (0 false alarms, no misses)
on the block-level inset/margin class it dominates, while zero-golden is the only mode that detects the decoration/styling-asymmetry class, at the price of a \texttt{border-inline-end}
false positive. Two implications follow. Because the differential carries recall---11 of the 13 real detections, with no
false alarm on any patched version---it is the natural pre-merge CI gate wherever both directions can be rendered; the
cheaper single-render zero-golden pass (median 55 vs.\ 110\,ms,
but recall-limited and prone to end-side false positives) is
best reserved as a lightweight audit for deployed or third-party builds that cannot be re-rendered LTR. And the eight-year persistence of physical directional CSS (Section~\ref{sec:history})
suggests this defect class is a maintenance constant of long-lived design systems, not a transient migration cost. Future
work: enlarge the held-out benchmark, infer directional-icon
annotations, cover more scripts and composed pages, and
add multi-node counterfactual repair evaluated against an
\emph{independent} oracle.

\section*{Acknowledgment}

\AcknowledgmentText

\end{document}